\documentclass[a4paper,prb,aps,twocolumn,groupedaddress,showpacs,showkeys]{revtex4}
\usepackage{graphicx,latexsym}
\begin{document}

\title{Possible effect of collective modes in zero magnetic field transport in an electron-hole bilayer}

\author{A.F. Croxall}\author{K. Das Gupta}\email{kd241@cam.ac.uk}\author{C.A. Nicoll}
\author{H.E. Beere} \author{I. Farrer} \author{D.A. Ritchie} \author{M. Pepper}\thanks{Present address: Department of Electronic and Electrical Engineering,University College, London, WC1E7JE}
\affiliation{Cavendish Laboratory, University of Cambridge, J.J. Thomson Avenue, Cambridge CB3 0HE, UK.}

\begin{abstract}
We report single layer resistivities of 2-dimensional electron and hole gases in an electron-hole
bilayer with a 10nm barrier. In a regime where the interlayer interaction is
stronger than the intralayer interaction, we find that an insulating state
($d\rho/dT < 0$) emerges at $T\sim1.5{\rm K}$ or lower,  when both
the layers are simultaneously present. This happens deep in the $\lq\lq$metallic" regime, even in layers with
$k_{F}l>500$, thus making conventional mechanisms of localisation due to
disorder improbable.  We suggest that this insulating state may be due to a
charge density wave phase, as has been expected in electron-hole bilayers from the Singwi-Tosi-Land-Sj\"{o}lander approximation based calculations of L. Liu {\it et al} [{\em Phys. Rev. B}, {\bf 53}, 7923 (1996)].  Our results are also in qualitative agreement with recent Path-Integral-Monte-Carlo simulations of a two component plasma in the low temperature regime [ P. Ludwig {\it et al}. {\em Contrib. Plasma Physics} {\bf 47}, No. 4-5, 335 (2007)]

\end{abstract}

 \maketitle
\date{\today}
\section{Introduction}
The  total energy of a system of electrons can be thought of as the sum total
of the kinetic energies of the free particles and their potential energies due
to mutual Coulomb interaction. The relative importance of the two contributions
is measured by the parameter $r_s=E_{ee}/E_{f}$ (where $E_{ee}=e^2\sqrt{(\pi N)}/{4\pi\kappa\epsilon_0}$
and $E_{f}=\pi\hbar^{2}N/m_{eff}$  in 2-dimensions, with N electrons per unit area). $r_{s}$ is a crucial parameter
  (apart from disorder) that governs the behaviour of the system to a large extent. For example it is known
  that a system of electrons or holes would be $\lq\lq$gas like" at $r_{s}\approx 1$, $\lq\lq$liquid like" at
   $r_{s}\approx 10$  and possibly a solid $\lq\lq$Wigner crystal" at $r_{s}\approx 100$, provided disorder driven
   localisation does not dominate and pre-empt the observable effects of interaction. Confining a large number
   of particles in a small area makes the inter-particle spacing small and hence the Coulomb repulsion large,
   but the kinetic energy of the particles increases even faster - making $r_{s}$ smaller. This somewhat
    counter-intuitive fact is a straightforward consequence of Fermi statistics and is true  in all  dimensions.\\
Let us now consider two parallel layers of electrons or holes with $10^{11}
{\rm cm}^{-2}$ electrons in each. As they are brought closer to each other the
particles in one layer not only interact with others in the same layer but also
with those in the other layer. The inter-particle spacing in the same layer
stays fixed and is about 30nm.  It is possible (though highly non-trivial) to
make the distance between the two layers about 10nm without their wavefunctions
beginning to overlap. 10nm is approximately the excitonic Bohr radius in
Gallium Arsenide (GaAs) and is an important length scale. We thus get an
electron to $\lq\lq$see" another electron (or hole) only 10nm away, without
paying the kinetic (Fermi) energy cost, because the two layers continue to be
two separate Fermi systems. 
As a consequence bilayer systems can
give rise to interaction-driven phases that are not possible in single layers.
The case of attractive interaction (electron-hole) is more interesting.
A rich phase diagram is anticipated  for the ground state of  a spatially separated electron-hole bilayer (EHBL) that includes a
superfluid\cite{lozovik,vignale,conti,hu,balatsky},  charge density waves (CDW), Wigner crystals (WC)\cite{liu_prb,depalo,moudgil}, an excitonic supersolid\cite{joglekar1} and a possible crossover from a
Bose-condensate to Bardeen-Cooper-Schreiffer type state\cite{pieri}. Recent  techniques for making independent ohmic contacts to 2-dimensional electron gases (2DEG) and 2-dimensional hole gases (2DHG)
 10-20nm apart have enabled transport measurements down to millikelvin
temperatures\cite{keogh,seamons1,croxalljap}. This  has  greatly extended the range of densities and temperatures  that could be   explored in the first attempt to measure transport in EHBLs \cite{sivan}.\\
Quite naturally the initial
experiments were focussed on measuring the Coulomb drag in these
bilayers\cite{croxallprl,seamons2}.  These results strongly suggest the emergence of a non Fermi-liquid phase at temperatures below
$T\sim1K$.  The scattering rate between the
electrons and holes (that is measured in Coulomb drag experiments) is not determined  by the bare Coulomb potential, as the inter-particle interaction is  screened and it is the dielectric
function $\epsilon(q,\omega)$, in the presence of interactions (and possible
pairing fluctuations\cite{hu} ) that determines the measured Coulomb
drag. On the other hand the dominant contribution to the single layer
resistance comes from the scattering due to residual charged background
impurities/traps and the remote ionised dopants (if any). The bare Coulomb
potential due to these is screened by the same dielectric function. Thus it is
worth asking the question: Does the interlayer interaction also bring about some distinct feature in the single layer resistances? This is the central question we investigate experimentally in this paper.
A connection between drag and single layer resistance was earlier investigated
in the context of in-plane magnetoresistance and magnetodrag in hole gases \cite{dassarma2dhgdrag}. There is however a difference between the
screening of interlayer electron-hole interactions and impurity potentials. The
electrons and holes are mobile and  the impurities are static. The potentials due to electrons/holes change with time and one needs to consider  dynamic
screening.  On the other hand static screening
$\epsilon(q,\omega=0)$ is sufficient to account for the effect of ionised
impurities. Thus the Coulomb drag is sensitive  to more features in the
screening function (such as plasmon modes). Also, it is easy to imagine
situations where the Coulomb drag and single layer resistivities would have
opposite temperature dependences. For example, if the layers are individually
insulating,  then the resistivity of the  layer
will increase as the temperature is decreased, but the Coulomb drag will go to
zero monotonically at low temperatures. While we are aware of these
distinctions between drag and single layer resistance, we find striking  features in the temperature dependence of
the single layer resistivities   of both layers. These features are
unambiguously attributable to the presence of the other layer and
hence electron-hole interaction only.\\
\section{Devices and experimental methods }
In our experiments the electron and hole densities ($n,p$)were varied between
$4\times10^{10}{\rm cm}^{-2}$ and $1.6\times10^{11}{\rm cm}^{-2}$, separated by a 10nm Al$_{0.9}$Ga$_{0.1}$As barrier. In this regime the inter-layer interaction is comparable to or even stronger than the intra-layer interaction.
Several important length scales in the problem are very close to
each other. Considering a typical density  of $1\times10^{11}{\rm cm}^{-2}$,
the inverse Fermi wavevector is $k_{F}^{-1}=1/\sqrt{2{\pi}n}\approx$13nm. The
hole quantum well is 20nm wide. The peak-to-peak distance of the
wavefunctions is $d\approx25$nm. For the lowest density
($n=p=4\times10^{10}{\rm cm}^{-2}$), we have $k_{F}d=1.2$. At the same time,
the (conduction band) excitonic Bohr radius in GaAs is $a_{B}\approx$10nm.
The interaction parameter  for the electrons, $r_{s}$ (ratio of the interparticle Coulomb energy to the Fermi energy) at the lowest density is 2.8 and  is $\approx$12 for the holes. Thus, we reach a regime $d\sim r_{s}a_{B}$ where excitonic phases \cite{lozovik,depalo}and density modulated phases\cite{liu_prb,moudgil} have been predicted.
The Molecular Beam Epitaxy growth, fabrication process and operation of these devices have been
described in detail earlier\cite{keogh,croxalljap}, so we do not describe them
here. Two samples (Device D \& E) fabricated from the same wafer (ID:A4268)
were used in this study and gave very similar results. (A third sample also gave similar results.) Coulomb drag data from device D was reported earlier\cite{croxallprl}.\\

 The complexity of these samples, at present, stands in the way of integrating coplaner waveguides and performing high frequency ($\sim$ 100 MHz) experiments to probe the $\lq\lq$pinning modes" of  density modulated phases. Such experiments have been done on bilayer hole gases, for example \cite{wang_prl}, at the cost of sacrificing independent contacts and independent tuning of the layer densities. These experiments, if they can be done on EHBLs in future, would certainly provide valuable data. However we show that even dc transport in EHBLs (with its limitations), yields results strongly suggestive of the appearance of CDW or WC phases in EHBLs.

\section{Results}
\begin{figure}[t]
\includegraphics[width=8.6cm,clip]{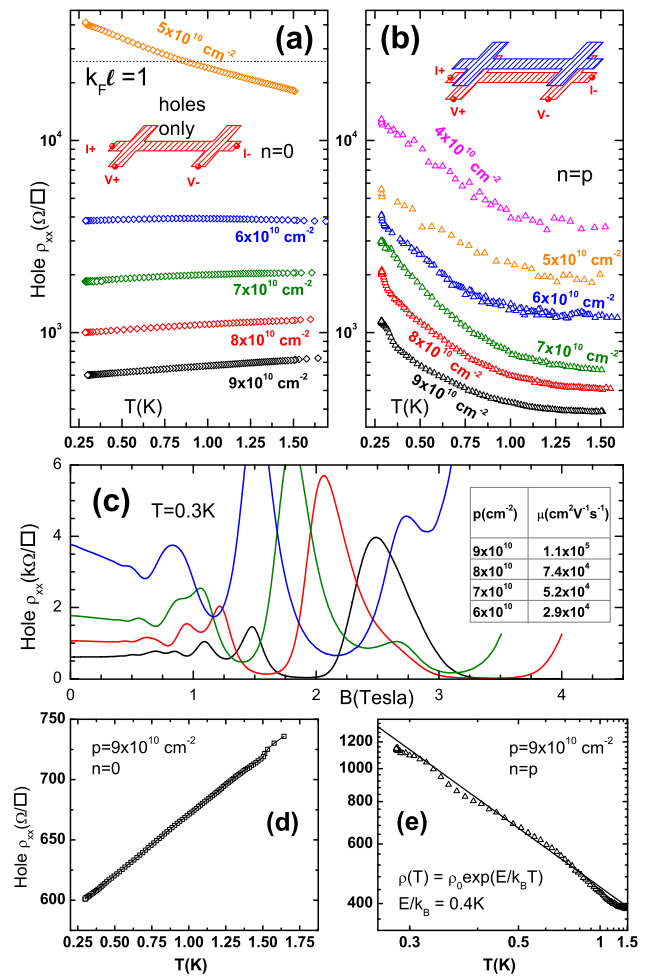}
\caption{\label{hrxxn0np}(Color online)Data from device E, measured in a pumped
He$^{3}$ cryostat. (a) $\rho_{xx}(T)$ of the hole layer, when no electrons
are present. (b) $\rho_{xx}(T)$ of the hole layer with electrons
present. Exactly the same current and voltage contacts, measuring current ($\sim 10{\rm nA}$) and frequency (7Hz) were used for consistency. (c) Shubnikov de-Haas data corresponding to the traces in (a) showing  normal 2DHG behaviour. The lowest trace corresponds to $p=9\times10^{10}{\rm cm}^{2}$. (d) "Metallic" behaviour of the holes at
$p=9\times10^{10}{\rm cm}^{2}$, with $n=0$. (e)Near exponential rise of
$\rho_{xx}(T)$, at $p=9\times10^{10}{\rm cm}^{2}$ with $n=p$. Note that the figure is plotted in a log vs $\frac{1}{T}$ scale.}
\end{figure}
We first show the temperature dependence of the resistivity of the holes, when
no electrons are present (Fig. \ref{hrxxn0np}a).  This was
achieved by keeping the interlayer bias voltage below the threshold
($V_{eh}\approx1.55{\rm V}$) for accumulation of electrons. The
density of the holes was controlled by using the  backgate. The data
shows features typical of numerous studies on Si-MOSFETS and GaAs based single layer 2DEGs or 2DHGs in the context of the 2D Metal-Insulator transition (MIT) problem. At low densities ($p{\leq}5\times10^{10}{\rm cm}^{-2}$) the traces are clearly insulating ($d\rho/dT < 0$), and turn metallic ($d\rho/dT > 0$) at higher densities. The crossover happens as expected near $\rho_{xx}\approx h/e^2$ or equivalently when $k_{F} l\approx1$. The central result of this paper is the strikingly different behaviour of the 2DHG when it sees the 2DEG.
 Fig. \ref{hrxxn0np}b, where we plot the hole resistivity for
several equal electron and hole densities, shows that even the traces which
were metallic in Fig. \ref{hrxxn0np}a ({\it e.g.} $p=9\times10^{10}{\rm cm}^{-2}$ in Fig. \ref{hrxxn0np}c) turn very clearly
insulating by $T\sim1K$ even at $k_{F}l > 50$. We will show later that $n=p$ is not essential for this insulating behaviour to appear.
Simple arguments based on localisation due to background
impurities/dopants/defects cannot account for this because the 2DHG (without
the 2DEG) would  see exactly the same charged background impurities. This
is the key point that allows us to attribute this effect to electron-hole
interactions. In fact if we look at the high temperature end of the traces
($T\approx1.5{\rm K}$) we find that the resistivity of the 2DHG actually
decreases significantly as the electrons are introduced. For the $p=6\times10^{10}{\rm cm}^{-2}$ trace,
$\rho_{xx}(n=0) = 3808\Omega/\Box$, whereas $\rho_{xx}(n=p) =
1203\Omega/\Box$ at $T=1.5{\rm K}$ (compare Fig. \ref{hrxxn0np}(a) and \ref{hrxxn0np}(b)). This is a very significant (threefold)
increase in mobility and consistent with the findings of Morath {\it et al} in
EHBLs with  30nm barriers\cite{morath1}. The effect is much stronger here, most
likely because of the much thinner barrier used in this study.  The increase in
mobility at the high temperature end is most likely brought about by the extra
screening of charged impurities due to  the 2DEG and the $\lq\lq$squeezing"
of the hole wavefunction itself. The Coulomb drag was also measured for both
the samples (data not shown) and at $T=1.5{\rm K}$ the values obtained are
consistent with a $T^2$ behaviour.  Below $T\approx1{\rm K}$ both samples show a  strong deviation from
$T^2$ behaviour, as reported recently by us and Seamons {\it et al}\cite{croxallprl,seamons2}.
(The $T^2$ behaviour can be well understood within Fermi-liquid theories, but the strong non-monotonic deviations cannot be.)
But a simple Mathiessen's rule based
addition of two contributions due to Coulomb drag and impurity scattering cannot explain an {\em increase} in mobility. At the same time (for $n$=$p$=$9{\times}10^{10}{\rm cm}^{-2}$) the measured Coulomb drag at $T<1{\rm K}$ is much smaller than the change in single layer resistivity (of the 2DHG) brought about by the presence of the 2DEG. Simple (incoherent) addition of various scattering rates no longer appears to be able to account for the observed magnitude of the effect.

\begin{figure}[t]
\includegraphics[width=8.6cm,clip]{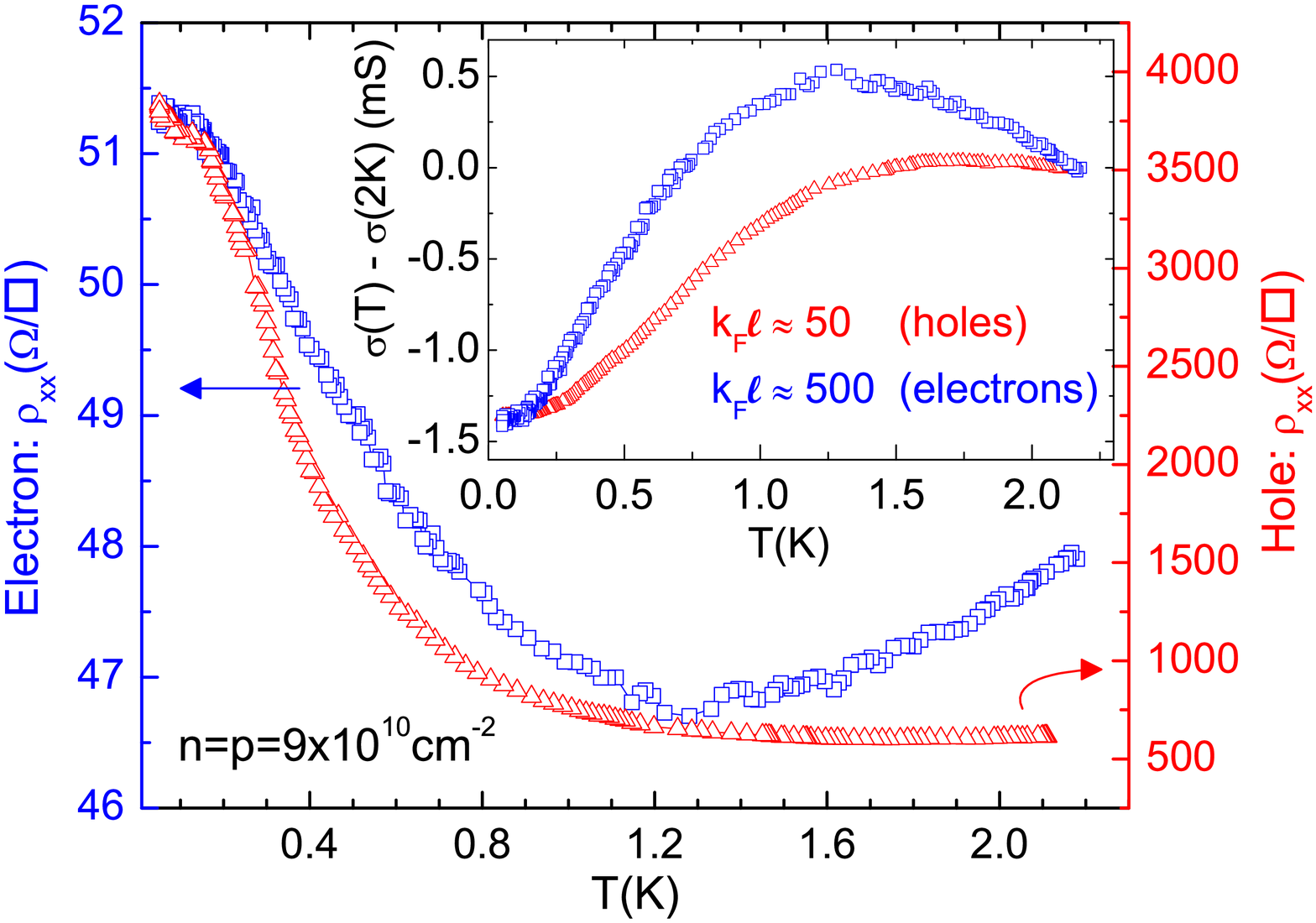}
\caption{\label{np9e10ehrxx}(Color online) A comparison of the electron(blue) and hole (red) resistivity data from device D, measured in a
dilution refrigerator with base temperature of $\approx50{\rm mK}$.  At this density were $\mu_{e}=1.5\times10^{6}{\rm
cm}^{2}{\rm V}^{-1}\rm{s}^{-1}$ \& $\mu_{h}=1.1\times10^{5}{\rm cm}^{2}{\rm
V}^{-1}\rm{s}^{-1}$ (at $T\approx1.5{\rm K}$).
The inset shows the amount of conductivity lost by the layers as a function of
temperature.}
\end{figure}

\begin{figure}[t]
\includegraphics[width=8.6cm,clip]{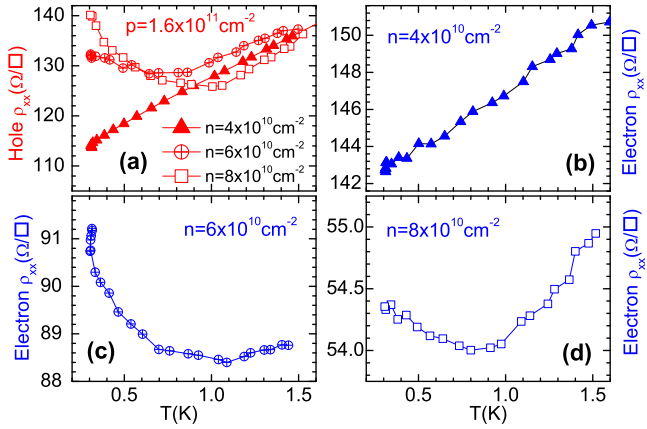}
\caption{\label{p16e10nvaries} Data from device E, measured in a pumped He$^3$
cryostat. The hole density was kept fixed at $p=1.6\times10^{11}{\rm cm}^{-2}$
while the electron density was varied in steps n=$4,6,8\times10^{10}{\rm
cm}^{-2}$. (a) The $\rho_{xx}(T)$ for holes is $\lq\lq$metallic" when the number of
electrons is  $n=4\times10^{10}{\rm cm}^{-2}$ on the other side of the barrier.
As the electron density is increased, the $\rho_{xx}(T)$ of the holes becomes insulating. (b), (c) \& (d) show the $\rho_{xx}(T)$ of the electrons. Note
that the electrons are metallic at low density and insulating at the two higher
densities, $k_{F}l\approx170$ or higher for all the traces.}
\end{figure}

If we now consider the behaviour of the electron layer, at
$n=p=9\times10^{10}{\rm cm}^{-2}$, the situation appears even more striking
because the insulating behaviour can be easily seen at $k_{F}l > 500$. Note
that (in Fig. \ref{np9e10ehrxx}) although the electron and hole densities are
same, the single layer resistances differ by nearly a factor of 10 at $T=1.5K$
and by a factor of 80 at $T\approx50{\rm mK}$. However the amount of
conductivity ($\Delta\sigma_{xx}$) lost by the layers as the temperature is reduced below 2K, does not differ by more
than a factor of 2.  $\Delta\sigma_{xx}$ between 2K and $\approx50{\rm mK}$, is
of the order of 2m$\mho$, which places it much beyond what weak localisation
based effects ($\Delta\sigma \sim e^{2}/h\approx40\mu\mho$) can account for
\cite{bergman}. In Fig. \ref{p16e10nvaries} we show how $d\rho/dT$ of both the
electron and hole layers below $\sim0.7{\rm K}$ turn from $\lq\lq$metallic" to
$\lq\lq$insulating" as the carrier density of the electron layer is {\em
increased} in steps, while keeping the hole density ($p=1.6\times10^{11}{\rm
cm}^{-2}$) constant.  We chose this density, such that the sheet resistivities
(at $T\sim1.5{\rm K}$) could be made nearly same for both the layers. The
insulating state gradually evolves from a $\lq\lq$metallic" state as the holes
see more electrons. As far as we can say, the insulating state appears
simultaneously in both layers. However,  $n=p$ does not appear to play a
particularly special role in the problem, as has been found \cite{croxallprl,seamons2} in the context of Coulomb drag in these bilayers.\\

\section{Discussion}
To understand what might be causing this we organise our discussion in  three steps. First we have
already mentioned several points which show that the physics of single layer 2DHG or 2DEG cannot
explain the results. Next we  consider whether  any inhomogeneity caused by the large
voltage bias across the 10nm barrier can affect our results. Finally we discuss the possibility of
the emergence of collective modes (like CDWs) in an EHBL and refer to  theoretical works that
suggest that such modes are more likely in  EHBLs than  2$\times$2DEGs or  2$\times$2DHGs.\\
Let us now consider the possible effect of some variation in thickness of the barrier,
resulting in spatial variation of the interlayer capacitance. The question here is: In our devices
does this lead to  coexisting low density (insulating) regions and high density (metallic) regions?
If we assume a variation of a monolayer ($\approx$0.5nm), this would amount to a maximum of 5\% fluctuation.
The corresponding density fluctuations would not (for example) be able to force  regions
with $p=9\times10^{10}{\rm cm}^{-2}$ and  regions with $p{\approx}5\times10^{10}{\rm cm}^{-2}$
(which is where we find the 2D MIT in the 2DHG when no 2DEG is present) in the same sample to coexist.
Unless the inhomogeneity is so strong that the low density regions become strongly insulating,
it cannot be the driving factor. If the 2DHG did become strongly inhomogeneous, it would have
increased the sheet resistance at high temperatures as well. As long as the underlying idea is
based on the general (single layer) picture of insulating behaviour above $h/e^2$ and metallic
behaviour below this resistance, it is not possible to explain the results. A fundamentally
different mechanism that can generate an insulating state at $\rho_{xx}<<h/e^{2}$ must be sought here.\\
We look for an approach that specifically takes into account that in a bilayer
there can be collective modes  that have no analogue in single layer situation.
Collective modes are zeros of the  dielectric function $\epsilon(q,\omega)$
where a system can develop density modulations without an external
perturbation. For a bilayer $\epsilon(q,\omega)$ is a $2{\times}2$ matrix,
whose determinant would be zero\cite{dassarma_madhukar, hu_plasma}.
Equivalently, the charge susceptibility of the system would diverge at these
points. If such a mode occurs at $\omega$=0, it can be a CDW or  WC phase,
depending on the wavevector at which the divergence occurs. In a bilayer an
electron  $\lq\lq$sees" another electron (or hole) only 10nm away, without
paying a kinetic (Fermi) energy cost. As a consequence bilayer systems can give
rise to interaction-driven phases more easily than single layers. In fact,
existing calculations\cite{liu_prb,moudgil} of the bilayers  for particle
densities corresponding to $r_s \sim 1-10$ mention  the possibility that a
divergence in the eigenvalues of the bilayer $\epsilon(q,\omega=0)$ matrix is
easier to get in an EHBL compared to 2$\times$2DEGs or 2$\times$2DHGs. This
point needs  attention because experimental data from 2$\times$2DEGs or
2$\times$2DHGs with a barrier separation of 10nm and densities as low as
$1{\times}10^{10}{\rm cm}^{-2}$ have been available  for some
time\cite{kellog,tutuc}. These samples had similar electron mobilities as the
EHBLs reported here, but higher hole mobilities. The principal interest with
these devices has been focussed on the $\nu=\frac{1}{2} + \frac{1}{2}$ bilayer Quantum Hall
state, but  interestingly  no collective mode at B=0 has been
reported in these\cite{ho_prb}. Following Liu {\it et al},\cite{liu_prb} we consider the expression for eigenvalues of
the susceptibility matrix, one of which must diverge to support a
CDW.
\mbox{$\chi_{\pm}(q)= \frac{2}{\chi_{e}^{-1} + \chi_{h}^{-1} \pm \sqrt{(\chi_{e}^{-1} -\chi_{h}^{-1})^{2} + 4\{(1-G_{eh}(q))V_{eh}(q)\}^{2}}}$}.
 $\chi_{e,h}$ are the single layer responses to the external potential and
$V_{eh}(q)=-\frac{2{\pi}e^{2}}{q}\exp(-qd)$ is the Fourier transform of the
interlayer Coulomb potential. The crucial role is played by the local field
correction $G_{eh}(q)$, often calculated using the
Singwi-Tosi-Land-Sj\"{o}lander (STLS) approximation\cite{stlspaper} that relates the local field to the structure factor.  The
interlayer local field correction is larger in magnitude in case of an
attractive potential\cite{liu_prb,moudgil}. Due to this a divergence of the
in-phase mode in an EHBL is easier to get than the
corresponding divergence in the out-of-phase mode in a 2$\times$2DEG or a 2$\times$2DHG. At the same time the strong peak in $\chi(q)$, (that signals the formation of a CDW state) has been found to be strongly temperature dependent, in  numerical studies \cite{liu_physicab}. The range of temperatures at which the rapid increase in resistance is seen is consistent with the expectation that it should be at a temperature smaller than the Fermi temperature of the layer with heavier mass ($T_F\approx 5{\rm K}$ for holes, at $p=9\times{10}^{10}{\rm cm}^{-2}$).\\

Path Integral Monte Carlo simulation of a spatially separated 2-component plasma has been carried out by Ludwig {\it et al}\cite{ludwig} for parameters relevant to GaAs-AlGaAs at low temperatures. The real space distribution of the holes (in presence of electrons) started developing density modulations, while the lighter electron liquid remained in a much more homogeneous state. Such a situation is quite consistent with our observations of an insulating behaviour of the holes, and a less pronounced (but clearly discernible) effect on the electrons. A  density modulation (at the same wavevector)  with larger amplitude in the hole layer and smaller amplitude in the electron layer is also supported by the STLS calculations \cite{moudgil}. \\

There is nothing in our observations that rules out a much awaited  excitonic state in  electron-hole bilayers.  The
interlayer pair correlation  in an excitonic (bound) state would involve a more
drastic modification of the structure factor and the local
field corrections, compared to that of a CDW state.  A CDW may well be a precursor to an excitonic state. \cite{liu_physicab}\\
Finally, we observe that a CDW may also be a possible
mechanism behind the temperature dependent enhancement in Coulomb drag reported
recently\cite{croxallprl,seamons2}. We consider a situation where the densities ($n$,$p$) in each layer
undergo an in-phase modulation $\delta$ with a long wavelength ({\em i.e.}
small $q/k_F$), such that the densities in successive half periods are
$n-\delta$ and $n+\delta$. We assume that we can evaluate the contribution to
drag over each {\em half cycle} and add the voltages in series. Over each half
cycle the contribution is still assumed to be proportional to $1/(np)^{3/2}$
, where $n$,$p$ are now the local densities\cite{jauho}. Thus the contribution
of two adjacent half cycles would be    $\propto
\frac{1}{(n+\delta)^{3/2}(p+\delta)^{3/2}} +
\frac{1}{(n-\delta)^{3/2}(p-\delta)^{3/2}}$. Keeping terms upto $\delta^2$ we
get the contribution from each CDW period to be $\propto
\frac{2}{(np)^{3/2}}\frac{1+
(9/4)\delta^2/np + (3/8)\delta^{2}(1/n^2 + 1/p^2)}{(1-\delta^2/n^2)^{3/2}(1-\delta^2/p^2)^{3/2}}$.  This quantity is clearly  larger than the unperturbed
contribution $\frac{2}{(np)^{3/2}}$, and it is clearly temperature dependent,
since the amplitude of a CDW ({\em i.e} $\delta$) can increase as  the temperature decreases. This
argument neither depends on the exact form of the power law nor the matching of the densities. The fact that it is  of order $\delta^2$ may have some
significance in explaining why the observed effect\cite{croxallprl,seamons2} is rather small and not a
large peak or divergence expected at matched densities from an excitonic phase.\\
Acknowledgements: The authors acknowledge useful discussions with A.R. Hamilton and D. Neilson and J. Waldie. The work was funded by EPSRC, U.K.

 \end{document}